# Silicon nanostructure cloak operating at optical frequencies

Lucas H. Gabrielli

*Nanophotonics Group, School of Electrical and Computer Engineering*

*209 Phillips Hall, Cornell University, Ithaca, NY 14853, USA*

*lhg28@cornell.edu*

Jaime Cardenas

*Nanophotonics Group, School of Electrical and Computer Engineering*

*209 Phillips Hall, Cornell University, Ithaca, NY 14853, USA*

*jc922@cornell.edu*

Carl B. Poitras

*Nanophotonics Group, School of Electrical and Computer Engineering*

*211 Phillips Hall, Cornell University, Ithaca, NY 14853, USA*

*cbp8@cornell.edu*

Michal Lipson*

*Nanophotonics Group, School of Electrical and Computer Engineering*

*214 Phillips Hall, Cornell University, Ithaca, NY 14853, USA*

*ml292@cornell.edu*

**The ability to render objects invisible using a cloak – not detectable by an external observer – for concealing objects has been a tantalizing goal[1-6]. Here, we demonstrate a cloak operating in the near infrared at a wavelength of 1550 nm. The cloak conceals a deformation on a flat reflecting surface, under which an object can be hidden. The device has an area of 225 µm$^2$ and hides a region of 1.6 µm$^2$. It is composed of nanometre size silicon structures with spatially varying densities across the cloak. The density variation is defined using transformation optics to define the effective index distribution of the cloak.**

The prospect of optical cloaking has recently become a topic of considerable interest. Through the use of transformation optics[7-11], in which a coordinate transformation is applied to Maxwell's equations, several designs for such a device were created[12-23]. These designs are based on the idea of manipulating the structure of the cloaking medium so that the trajectory of light after interacting with the cloak is the same as that in an empty medium, without the cloak or the object underneath. The external observer is therefore unaware of the presence of the cloak and the object. Such cloaks were recently experimentally demonstrated in the microwave regime using metamaterial structures with feature sizes in the millimetre to centimetre scale[24,25]. Pushing this technology to the optical regime would greatly increase the potential application. This requires, however, nanometre control of the cloaking structure.

Here we demonstrate a cloak in the optical domain operating at 1550 nm using sub-wavelength scale dielectric structures. We experimentally demonstrate an optical invisibility cloak that hides an object "under a carpet" with the help of a reflective surface. As shown in the sketch of Fig. 1, when an external observer looks at a deformed mirror, it detects the deformation in the reflected image (Figs. 1a, b). Following the theoretical work by Pendry and co-authors[20] and Leonhard and co-authors[11], we designed and fabricated a cloaking device at optical frequencies capable of reshaping this reflected image and providing the observer with the illusion of looking at a plane mirror (Fig. 1c). One could then hide objects under those deformations without revealing their existence. Li has also demonstrated a similar device using a different fabrication technique[26].

The cloaking device has a triangular shape with 225 µm$^2$ of area, and is composed of a spatially varying density of sub-wavelength 50 nm diameter silicon posts embedded in an SiO$_2$ medium. The reflective surface consists of a distributed Bragg reflector (DBR) with a deformation that covers the 1.6 µm$^2$ cloaked region. The distribution of posts induces a variation of the effective index of refraction across the surface through the relation $\varepsilon_{eff} = \rho_{SiO2}\varepsilon_{SiO2} + \rho_{Si}\varepsilon_{Si}$, where ρ is the volumetric fraction and ε is the effective dielectric constant of each material[27]. The DBR consists of alternating regions of SiO$_2$ and crystalline silicon. The simulated reflectivity for the ten-period DBR used is larger than 0.999. We fabricate the invisibility cloak in a silicon-on-insulator (SOI) wafer. An etching mask, consisting of a layer of 160 nm of XR-1541®, is patterned by E-beam lithography, and the 250 nm top silicon layer is etched using a standard Cl$_2$ inductively coupled plasma process. We then clad the device with SiO$_2$. Scanning electron microscope images of the fabricated device before the deposition of the SiO$_2$ are shown in Fig. 2. A 450 nm wide silicon waveguide with a tapered end reaches the device at the mid-point of the edge on the *y* axis, and is used for input. We use the waveguide to input light in the device in order to ensure that all of the input light is incident on the deformation and not on the plane DBR reflector, therefore maximizing the effect of the deformation. Note, however, that the design based on transformation optics does not introduce any constraints on the wave fronts applied to the device[20], which means that the cloaking medium operates at all angles of incidence where the reflectance of the DBR is high enough. In Fig. 2b one can see the reduced density of silicon posts in the low effective index region of the cloak, as well as some of the silicon sections that compose the DBR.

The spatial distribution of the 50 nm diameter posts, i.e., the effective refractive index distribution of the cloak, is determined by defining a transformation of coordinates from the perfect triangle in Fig. 2a to one with a Gaussian-shaped deformation along its hypotenuse (behind which an object could in principle be hidden) for TM polarized fields (major component of the electric field perpendicular to the device). In order to minimize the anisotropy in the medium, the transformation of coordinates is realized by the minimization of the Modified-Liao functional[20, 28, 29] with slipping boundary conditions. The resulting effective index distribution has an anisotropy factor of 1.02 with index values ranging from 1.45 to 2.42, between the index of $SiO_2$ and that of crystalline silicon, enabling the fabrication of the device using standard silicon processes. The complete effective refractive index distribution is shown in Fig. 3. One can see the triangular shape of the device with the deformation along its hypotenuse. The highest and lowest effective refractive index regions are located around the deformation, and the background index value of the remaining cloaking region is 1.65. The final profile of the cloak contains almost no silicon in the low index regions; while in the high index regions it has the largest concentration of posts (see Fig. 2).

We simulate the propagation of light in the device using the finite difference time domain (FDTD) method and show that due to the presence of the cloak, the image of the light incident on the deformation, the region under which an object could be hidden, resembles the image of a wave that is propagating in a homogenous medium without the deformation. Figure 4a shows a simulation of light propagating through a homogenous background index of 1.65 and reflected by the DBR. Figure 4b shows the same simulation when the DBR is deformed, and Fig. 4c shows the simulation of light reflected by the deformed DBR (same as Fig. 4b), but now with the deformation covered by the cloak. By comparing Figs. 4a and 4b, one can clearly see that the presence of the deformed region can easily be observed in the output image of the device. Figure 4a, the plane DBR in a homogeneous medium, shows a uniform distribution of power along the bottom edge of the device (output) while the deformed mirror (Fig. 4b) presents a power gap as a result of the deformation, creating a power gap at the edge of the device. When the deformation is covered by the cloak (Fig. 4c), this power gap – a signature of the deformation on the DBR – disappears, and the output image resembles the case of light propagating through a homogeneous medium and reflected by the plane DBR (Fig. 4a).

We show experimentally, using an infrared camera, that the output of the light propagating through the cloak and incident on the deformation in the DBR mirror resembles the image from a plane mirror with no deformation. The device is tested by launching light with a wavelength of 1550 nm in the waveguide (shown in Fig. 2) and capturing the image at the edge of the device on the *xz* plane. The infrared camera used for the measurements was set to constant gain, contrast and brightness. The recorded images were filtered using a median filter, with a filter aperture of 15 pixels. Figure 5a shows the image of light reflected from a plane DBR in a homogeneous medium with index 1.65. In Fig. 5b one can see that the deformation of the reflector produces the power gap, as predicted from the simulation of the uncloaked deformed mirror. Figure 5c displays the image of light incident on the same deformed DBR, but now covered by the cloaking device. The power gap vanishes, and the image is similar to that reflected from the plane DBR and propagating in a homogeneous medium, as expected.

These results represent the experimental demonstration of an invisibility cloaking device at optical frequencies. The bandwidth and wavelength of operation of the device is limited by the bandwidth of operation of the distributed Bragg reflector and, for shorter wavelengths, silicon dispersion. This bandwidth is large, 400 nm, around a wavelength of 1550 nm due to the large index contrast between silicon and $SiO_2$. Such a cloak could in principle be reproduced over much larger domains, using techniques such as nanoimprinting, for example, enabling a wide variety of applications in defence, communications, and other industries. Note that in this paper we show how the trajectory of light can be manipulated around a region to render it invisible. Using transformation optics in a similar fashion to the one used in this paper, one could do the opposite - concentrate light in an area. This could be used for example for efficiently collecting sunlight in solar energy applications[30].

**Acknowledgements**

The authors would like to acknowledge the support of Cornell's Center for Nanoscale Systems (CNS), funded by the National Science Foundation. This work was performed in part at the Cornell Nanoscale Facility, a member of the National Nanotechnology Infrastructure Network, which is supported by the National Science Foundation.

**Author contributions**

L.H.G. designed and simulated the devices. L.H.G. and J.C. carried out the fabrication of the samples. L.H.G. and C.B.P. conducted the experiments. L.H.G., C.B.P. and M.L. designed the experiments and discussed their results and implications.

**Competing financial interests**

The authors declare that they have no competing financial interests.

**Figures:**

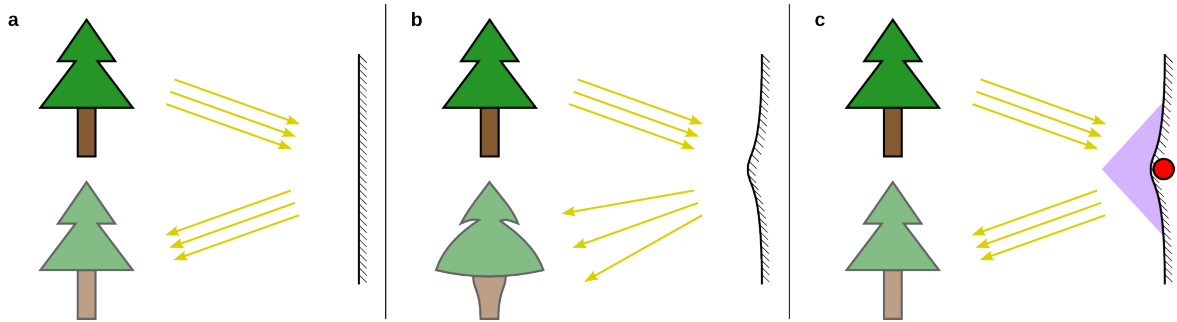

**Figure 1.** Cloaking principle of the fabricated device. (a) A planar mirror forms an image equal to the object reflected, but (b) when the mirror is deformed, the image is distorted, allowing an external observer to identify the deformation. (c) Our cloaking device – shaded area in front of the mirror – corrects the distortion in the image, so that the observer no longer identifies the deformation in the mirror, nor an object hidden behind the deformation.

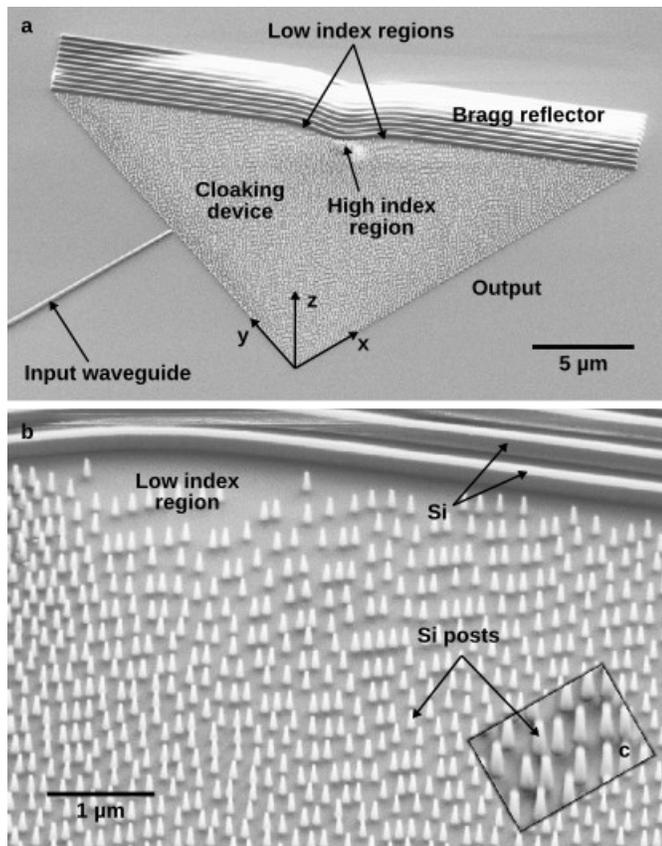

**Figure 2.** Scanning electron microscope images of the cloaking device. (a) Light is coupled to the device through the tapered input waveguide and reflected at the Bragg mirror towards the *xz* plane. (b) Silicon posts etched in the SOI wafer with varying density, inducing the effective index distribution of the cloak. (c) Zoom-in on some of the Si posts.

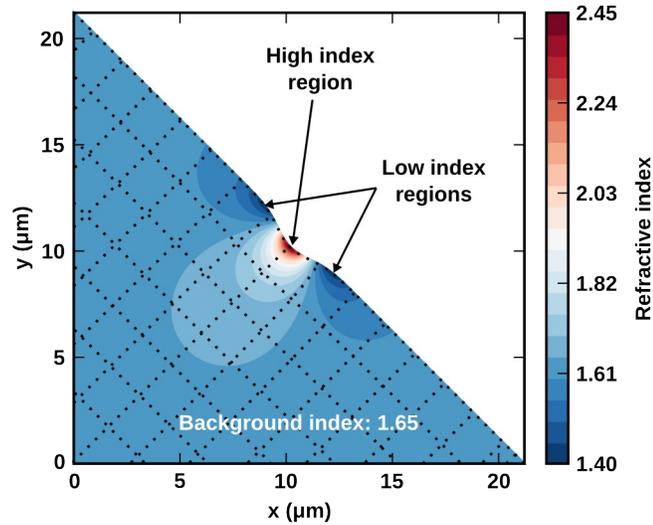

**Figure 3.** 2D spatial effective refractive index distribution of the cloaking device. The effective index values are between 1.45 and 2.42. The effective background index is 1.65. The dotted lines show the grid transformation.

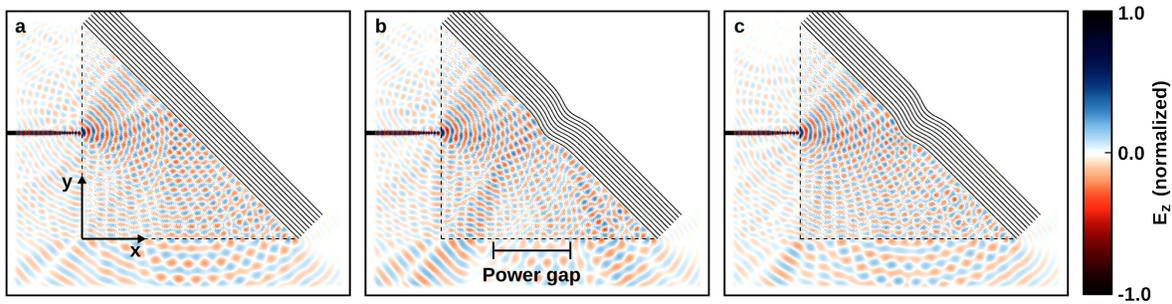

**Figure 4.** Simulations of the cloaking device. (a) Light at 1550 nm propagating through a homogenous background index of 1.65 and reflected by a plane DBR. (b) Same as (a), but with the DBR deformed. (c) Same as (b), but with the deformation covered by the cloak. The deformed mirror (b) presents a power gap as a result of the deformation, creating a power gap at the edge of the device. When the deformation is covered by the cloak (c) this power gap – a signature of the deformation on the DBR – disappears.

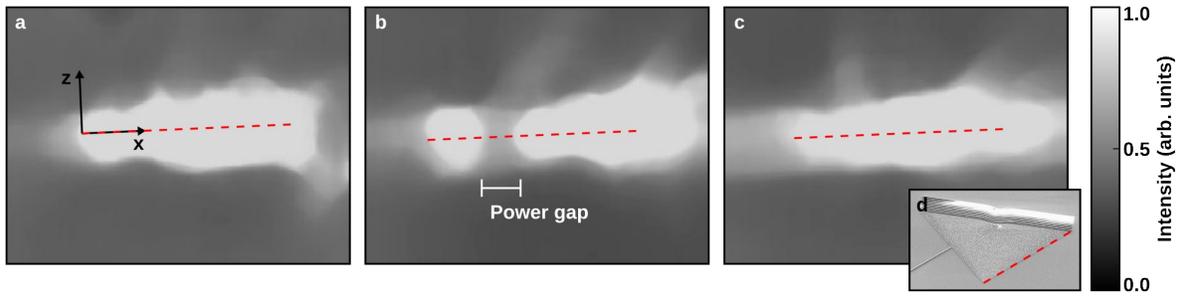

**Figure 5.** Output images from the fabricated devices, tested by launching light with a wavelength of 1550 nm in the waveguide and imaging the edge of the device (dashed lines). These experimental results correspond to the simulation predictions (Fig. 4). (a) Light reflected from plane DBR in a homogeneous medium (n=1.65). (b) Light reflected from deformed reflector without the cloaking device. (c) Light reflected from deformed DBR with the cloaking device, where the power gap vanishes resulting in an image similar to that reflected from the plane DBR (a), as expected. (d) Location of the output edge on the device.